\documentclass[pre,aps,onecolumn,superscriptaddress,showpacs,floatfix]{revtex4}
\usepackage{epsfig,graphicx,latexsym}
\usepackage{amssymb,amsmath}

\begin{document}

\pacs{05.10.Gg, 02.50.Ey}
\title{Approach to asymptotically diffusive behavior for Brownian particles in periodic potentials : extracting information from  transients.}
\author{David S. Dean}
\affiliation{Universit\'e de  Bordeaux and CNRS, Laboratoire Ondes et
Mati\`ere d'Aquitaine (LOMA), UMR 5798, F-33400 Talence, France}
\author{Gleb Oshanin}
\affiliation{Sorbonne Universit\'es, UPMC Univ Paris 06, UMR 7600, LPTMC, F-75005, Paris, France}
\affiliation{CNRS, UMR 7600, Laboratoire de Physique Th\'{e}orique de la Mati\`{e}re Condens\'{e}e, F-75005, Paris, France}
\begin{abstract}
A Langevin process diffusing in a periodic potential landscape has a time dependent diffusion constant which means that its average mean squared displacement (MSD)  only becomes linear at late times. The long time, or effective diffusion constant, can be estimated from the slope of a linear fit of the MSD at late times. Due to the cross over between a short time microscopic diffusion constant, which is independent of the potential, to the effective late time diffusion constant, a linear fit of the MSD will not in general pass through the origin and will have a non-zero constant term. Here we address how to compute the constant term and provide explicit results for Brownian particles in one dimension in periodic potentials.  We show that the constant is always positive and that at low temperatures it depends on the curvature of the minimum of the potential. For comparison we also consider the same question for the simpler problem of a symmetric continuous time random walk in discrete space. Here the constant can be positive or negative and can be used to determine the variance of the hopping time distribution.
\end{abstract}

\maketitle
\section{introduction}
Brownian motion is one of the most ubiquitous transport phenomena in Nature, thermally induced molecular collisions in a solvent impart momentum to small particles and they hence acquire a random  velocity \cite{satyabm}. In over damped systems, where inertia is overcome by viscous damping, this velocity  becomes rapidly uncorrelated in time. In this case the particle's velocity becomes an effective white noise and its position is what is mathematically known as pure Brownian motion. The diffusion equation describing the probability density (PDF) of pure Brownian motion in an isotropic medium in $d$ dimensions is 
\begin{equation}
{\partial p({\bf x},t)\over \partial t} = \kappa\nabla^2 p({\bf x},t).
\end{equation} 
The term $\kappa$ above is the short time or microscopic diffusion constant.
A key quantity which can be extracted from single particle tracking is the mean squared displacement 
(MSD) whose average is given by
\begin{equation}
\langle ({\bf X}_t-{\bf X}_0)^2\rangle = 2d\kappa(t) t.
\end{equation}
Here $\kappa(t)$ is the effective time dependent diffusion constant. The late time or effective diffusion constant is  defined via the limit
\begin{equation}
\kappa_e = \lim_{t\to\infty}\kappa(t).
\end{equation}
In the case of pure  Brownian motion we have $\kappa(t) = \kappa=\kappa_e$.

In Nature Brownian particles also interact with their environment via long range interactions that can be generated by electrostatic and other interactions. As well as the noise leading to Brownian motion, the particle also feels a  force, or drift, due to this potential. For instant a particle diffusing in a cellular cytoplasm will locally diffuse, but also interact via steric interactions with the cells organelles. The local or short time diffusion constant is solely determined by the nature  of the solvent, however at late times the diffusion constant will be modified due to the potential felt by the particle.  In the presence of drift due to
an external force it is easy to see that $\kappa(0) = \kappa$, {\em i.e.} the small time limit of the effective diffusion constant is given by the microscopic diffusion constant. However at late times the diffusion constant will be modified by the potential acting on the particle. The value of $\kappa_e$ can be computed exactly in one dimension for any potential \cite{1d}. However, except for a particular case in two dimensions \cite{2dscat}, in higher dimensions no results are known. The problem of diffusion in a random potential has been extensively studied, in the cases where the diffusion constant exists one can use field theoretic methods such as the perturbation theory and the renormalization group to estimate it
for both Gaussian \cite{ddh,deem} and non-Gaussian potentials \cite{2dscat,scat,gauss2,nongauss}. A review of these results can be found in \cite{review}. More recently, due to  increasing interest in the  physics of non-equilibrium steady states, there have been many studies of diffusion in tilted periodic  potentials, which are periodic potentials plus a constant linear potential leading to an component with a constant applied force \cite{tilt}.

The presence of a potential acting upon a Brownian particle will cause the effective trajectory to be non-pure Brownian and non-Gaussian. The effective diffusion constant must switch over from the microscopic one to the effective one, meaning that $\kappa(t)$ will vary with time. It can be shown that $\kappa_e<\kappa$, essentially due to trapping in the local minima of the potential,  so $\kappa(t)$ must decrease with time toward its asymptotic value. Consequently the MSD is concave rather than linear as in the case of Brownian motion, for instance see Fig (\ref{msd}) which is for particles diffusing in a cosine potential. The concave form of the MSD is superficially reminiscent of the sub-diffusive behavior $\langle ({\bf X}(t)- {\bf X}(0))^2\rangle \sim t^\alpha$ where $\alpha <1$ \cite{bouge}. It is thus conceivable that a slow crossover from the microscopic 
diffusion constant to the effective one could be misconstrued as anomalous sub-diffusion if the trajectories are not observed over  a  time sufficient to attain the asymptotic linear diffusive regime of the  MSD. The nature of the cross over is thus important to understand, both for discerning between anomalous and ordinary diffusion and because the crossover may contain information about the potential that the particle is subjected to. In this paper we will address the question of how an ultimately diffusive system attains the diffusive regime. 

To start with, we consider the finite time correction for continuous time random walks (CTRW). The finite time corrections here are quite easy to compute but the results are nonetheless illuminating and somewhat surprising. Furthermore,  at a coarse grained level a Brownian particle diffusing in a periodic potential can be viewed as a  CTRW. We then use a Kubo like formula, first given in \cite{scat}, to analyze finite time corrections for diffusion in a periodic potential and give explicit results for the case of one dimension. Our  results are also verified with numerical simulations based on the direct integration of the Langevin equation for an ensemble of tracer particles.

\section{Continuous time random walks}

To start with we consider the simple problem of CTRW in one dimensions on the set of integers.  The waiting time at each site has the same distribution $p(\tau)$. After waiting for time $\tau_i$ at site $i$ the process hops to the left or right with probability $1/2$. In terms of the number of jumps $N(t)$ taken up till time $t$ the mean squared displacement is just given by
\begin{equation}
\langle X_t^2\rangle = N(t),
\end{equation}
where $N(t)$ is itself a random variable depending on the waiting times, and the average $\langle \cdot\rangle$ is over the random variables corresponding to a jump to the left or right. The full average is given by
\begin{equation}
\overline{\langle X_t^2\rangle} =\overline{N(t)},
\end{equation}
where the average $\overline{\cdot}$ indicates the average with respect to the waiting times. Now consider an exponentially distributed random time $T$ with rate $s$. Consider the random variable
$N(T)$, the number of jumps made before the time $T$ occurs. The probability that a single jump
occurs before time $T$ is given by $p_1= {\rm Prob}(T>\tau)$ which is given by
\begin{equation}
p_1=\int_0^\infty d\tau\int_0^\infty d\tau's\exp(-s\tau')p(\tau)\theta(\tau'-\tau) =\int_0^\infty d\tau\ p(\tau)\exp(-s\tau) =\tilde p(s),
\end{equation}
were $\theta$ is the heaviside function and $\tilde p(s)$ is the Laplace transform of $p(\tau)$. Now using the memoryless property of the 
exponential distribution we see that the probability that there are $n$ steps taken before $T$ is given by
\begin{equation}
P(n) = \tilde p(s)^n(1-\tilde p(s)),
\end{equation}
from which we find that 
\begin{equation}
\mathbb{E}(\overline N(T)) = {\tilde p(s)\over 1-\tilde p(s)}\label{lt1},
\end{equation}
where $\mathbb{E}(\cdot)$ denotes the average over $T$.  Recalling that the probability distribution function of $T$ 
is given by $\rho(t) = s\exp(-st)$ means  that Eq. (\ref{lt1}) can be written as
\begin{equation}
 {\widetilde{\overline N}(s)} = {1\over s} {{\tilde p(s)}\over 1-{\tilde p(s)}}. 
\end{equation}
Therefore
\begin{equation}
\overline{\langle X_t^2\rangle} = {\cal L}^{-1}\left\{ {1\over s} {{\tilde p(s)}\over 1-{\tilde p(s)}}\right\},
\end{equation}
where ${\cal L}^{-1}\{\cdot\}$ indicates the inverse Laplace transform with respect to parameter $s$.

The late time behavior of the  MSD can be extracted from the small $s$ behavior of the Laplace transform, where we can use
\begin{equation}
\tilde p(s) = \overline{ \exp(-s\tau)}\approx 1- s\overline \tau + {s^2\over 2} \overline \tau^2 ,
\end{equation}
in the case where the two first moments of the distribution for $\tau$ exist. This gives 
\begin{equation}
\overline{\langle X_t^2\rangle} = {\cal L}^{-1}\left\{{1\over s^2\overline \tau} +{1\over s} \left[{\overline{\tau^2}-2{\overline\tau}^2  \over2 s{\overline\tau}^2}\right]\right\} .
\end{equation}
Now inverting the Laplace transform gives
\begin{equation}
\overline{\langle X_t^2\rangle} \approx {t\over \overline \tau} +\left[{\overline{\tau^2}-2{\overline\tau}^2  \over  2{\overline\tau}^2}\right] .
\end{equation}
This means that asymptotically the MSD is simply a straight line, measuring the slope gives the value of
$2\kappa_e$ while the intercept with the $y$ axis gives $2C$. The  MSD is usually determined from a finite ensemble of trajectories and even from a single trajectory by slicing up the trajectory and using different points for the origin of the process. The resulting experimental curve can be fitted, for example using a least squares linear fit.

Immediately we see that the effective diffusion constant is given by
\begin{equation}
\kappa_e = {1\over 2\overline \tau},\label{eqdc}
\end{equation}
which is intuitively obvious to understand, and shows the necessity of a finite average of $\tau$ to 
see diffusive behavior. An interesting point to make here is that in experiments one may not necessarily be able to obtain the temporal resolution necessary to see individual jumps and thus measure the average trapping time $\overline \tau$, however this can be obtained from the diffusion constant via
Eq. (\ref{eqdc}).

The time dependent diffusion constant $\kappa(t)$  therefore behaves at late times as 
\begin{equation}
\kappa(t) = \kappa_e + {1\over t}\left[{\overline{\tau^2}-2{\overline\tau}^2  \over 4{\overline\tau}^2}\right].
\end{equation}
The late time correction to the diffusion constant can be positive or negative depending on the sign of 
\begin{equation}
C = {\overline{\tau^2}-2{\overline\tau}^2  \over 4 {\overline\tau}^2}.
\end{equation}
Usually in fitting procedures to estimate $\kappa_e$ the value of the constant $C$ is ignored. Here we see that it contains information about the variance of trapping times. Using the estimate for $\overline \tau$  from the fit of $\kappa_e$ enables one to obtain and estimate for the variance ${\rm var}(\tau)$ from the estimate of $C$. 

Despite the simplicity of the above calculation, there are a number of interesting features which emerge. In the case where $\tau$ has an exponential distribution $p(\tau)=\mu\exp(-\mu\tau)$ we find that $C=0$. However in this case we can find the full temporal dependence as
\begin{equation}
\tilde p(s) = {\mu\over \mu + s}
\end{equation}
which yields
\begin{equation}
\overline{\langle X_t^2\rangle} = {\cal L}^{-1} \left\{ {\mu \over s^2} \right\}= \mu t
\end{equation}
{\em i.e.} the asymptotic diffusive regime sets in immediately.

We also note that the term $C$ can be written as 
\begin{equation}
\label{var}
C={ \rm{var}(\tau) -{\overline \tau}^2\over 4{\overline\tau}^2} = {1 \over 4} (c_v^2 - 1)\,,
\end{equation}
where ${\rm var}(\tau)$ is the variance
 and $c_v = \sqrt{{\rm var}(\tau)}/\overline\tau$ is the corresponding coefficient of variation 
of the parental waiting-time distribution. 
Therefore if $\tau$ has a large variance (so that $c_v > 1$) the late time correction to $\kappa(t)$ tends to be positive, however  if the distribution is highly peaked (and $c_v <1$) then the correction will be negative. The finite time correction for  peaked distributions is easy to understand by considering the {\em deterministic} hopping case where $p(\tau)=\delta(\tau-\tau_0)$ up to an infinitesimal dispersion of the distribution about $\tau_0$. Working in terms of the average jump number we find that
\begin{equation}
 \overline{ N(t)}= {t\over \tau_0} - {1\over 2}.
\end{equation}
However we know that
\begin{equation}
N(t) = {t\over \tau_0} - R({t\over \tau_0}),
\end{equation}
where $R(u)$ is the non-integer  part of $u$. The temporal average value of of this term is
given by,
\begin{equation}
\overline {R({t\over \tau_0})} = \int_0^1 du u = {1\over 2},
\end{equation} 
which  trivially explains the result for very peaked distributions, the constant $C$ is essentially negative due to the rounding effect of a discrete random walk.

Another distribution which is interesting to analyze is the power law distribution given by 
\begin{equation}
\label{broad}
p(\tau) ={ \alpha \tau_0^\alpha\over \tau^{1+\alpha}},
\end{equation} 
with $\tau_0$ a cut-off timescale below which $p(\tau)=0$.
The above analysis above goes through when  $\overline{\tau^2}$ is finite, that is  when $\alpha>2$.
In this case we find
\begin{equation}
C = {-\alpha^2 + 2\alpha +1\over 4\alpha (\alpha-2)}= -{(\alpha -\sqrt{2}-1)(\alpha +\sqrt{2}-1)\over 4\alpha (\alpha-2)}.
\end{equation}
Therefore there is a window of values of $\alpha$ such that $2<\alpha<\sqrt{2}+1$ 
such that $C>0$, and for all other values of $\alpha > \sqrt{2}+1$ we have $C<0$.  

Now consider the case where $\alpha\in (1,2)$ that is to say the mean exists but the variance diverges. In this case we find that
\begin{equation}
\tilde p(s) = 1- s\overline{\tau} +{s^\alpha \tau_0^\alpha \alpha}\int_{s\tau_0}^\infty {\exp(-u)-1+u\over u^{\alpha+1}}.
\end{equation}
Now in the limit where $s\to 0$ we can approximate the above by
\begin{equation}
\tilde p(s) = 1- s\overline{\tau} +{s^\alpha\tau_0^\alpha \alpha}\int_0^\infty {\exp(-u)-1+u\over u^{\alpha+1}},
\end{equation}
as the integral on the left hand side converges around $u=0$ for $\alpha <2$. This gives, for late times,
\begin{equation}
\overline{\langle X_t^2\rangle} \approx {\cal  L}^{-1} \left\{ {1\over s^2 \overline \tau} +
{ \alpha H(\alpha) s^{\alpha-3} \tau_0^\alpha\over \overline{\tau}^2} \right\} = {t\over \overline \tau}
+ {\alpha H(\alpha) \tau_0^\alpha t^{2-\alpha}\over \overline{\tau}^2 \Gamma(3-\alpha)}
\end{equation} 
where
\begin{equation}
H(\alpha) = \int_0^\infty du \ {\exp(-u)-1+u\over u^{\alpha+1}}= \Gamma(-\alpha).
\end{equation}
This can then be simplified, using the factorial property of the Gamma function $\Gamma(z)=(z-1)\Gamma(z-1)$, to obtain
\begin{equation}
\overline{\langle X_t^2\rangle} \approx {\cal  L}^{-1} \left\{ {1\over s^2 \overline \tau} +
{ \alpha H(\alpha) s^{\alpha-3} \tau_0^\alpha\over \overline{\tau}^2} \right\} = {t\over \overline \tau}
+{ \tau_0^\alpha t^{2-\alpha}\over (\alpha-1)(2-\alpha) \overline{\tau}^2 }
\end{equation}
We see that as the diffusion tends towards the point where it becomes anomalous (sub-diffusive),  the finite time corrections become more important. This is thus an example of a censorship phenomenon for the transition between diffusive and sub-diffusive transport. As the transition is approached the finite time corrections to the  MSD decay more and more slowly and become of the order of the leading, diffusive, term. The effective diffusion constant close to the transition point thus becomes impossible to measure if 
one does not know how to include finite time corrections in the fitting procedure used to extract $\kappa_e$. A concrete example of this was demonstrated for the case of a particle diffusing in a potential given by the square of a Gaussian function \cite{gauss2}, a system which exhibits a transition from a normal to anormal diffusive regime. 

Consider now the case $\alpha = \sqrt{2} + 1$, where $C$ in Eq. \eqref{var} is equal to zero, and address the question of the thermalization dynamics for this {\it critical} value of $\alpha$. For $2 < \alpha < 3$, i.e. when the second moment of the waiting time distribution in Eq. \eqref{broad} exists, while the third one does not, we have
\begin{equation}
\tilde{p}(s) \approx 1 - \overline{\tau} s + \frac{\overline{\tau^2}}{2}  s^2 - \alpha\tau_0^{\alpha} h(\alpha) s^{\alpha},
\end{equation}
where
\begin{equation}
h(\alpha) = - \int^{\infty}_0 \frac{d u}{u^{\alpha + 1}} \left(e^{- u} -1 + u - \frac{u^2}{2}\right) = -\Gamma( - \alpha).
\end{equation}
Consequently, we have that for $2 < \alpha < 3$,
\begin{equation}
\langle \overline{X_t^2} \rangle \approx L^{-1}\left\{\frac{1}{\overline{\tau} s^2} \left(1 + \left(\frac{\overline{\tau^2}}{2 \overline{\tau}} - \overline{\tau}   \right) s - \frac{\alpha\tau_0^{\alpha} h(\alpha)}{\overline{\tau}} s^{\alpha - 1}  \right)\right\} \,.
\end{equation}
Upon inverting the Laplace transform we find that
\begin{equation}
\langle\overline{X_t^2}\rangle \approx {t\over \overline \tau} +\left[{\overline{\tau^2}-2{\overline\tau}^2  \over  2{\overline\tau}^2}\right] -{\tau_0^\alpha t^{2-\alpha}\over\overline \tau^2  (\alpha-1)(\alpha-2)}.
\end{equation}
The latter equation implies that in the critical case $\alpha = \sqrt{2} + 1$ the long-time relaxation proceeds
as
\begin{equation}
\kappa(t) \approx \frac{1}{\sqrt{2} (\sqrt{2} + 1) \tau_0 } \left(1 - \frac{\tau_0^{\sqrt{2}}}{t^{\sqrt{2}}}\right) \,,
\end{equation} 
i.e., the diffusion coefficient approaches its equilibrium value from below and as a power-law with the exponent $\sqrt{2}$. We note that a similar {\it singular} behavior - i.e., the change of the dynamical exponent characteristic of the thermalization kinetics - is not specific to the power-law distribution in Eq. \eqref{broad} but may show up for any underlying waiting-time distribution $p(\tau)$ in which, by tuning some parameters, one  manages to tune the coefficient of variation $\sqrt{{\rm var} (\tau)}/\overline{\tau}$ of the underlying distribution $p(\tau)$ to be equal to $1$.

\section{Brownian particles in periodic potentials}

Now having considered the finite time correction for continuous time random walks on discrete space, where we have seen that the spatial discreteness can lead to  a trivial rounding type error, let us consider a Langevin process diffusing in a periodic potential.

Here we examine a locally Brownian particle whose probability density function (PDF) obeys the Fokker-Planck equation
\begin{equation}
{\partial \over \partial t}p({\bf x},t) = \kappa\nabla\cdot\left(\nabla p({\bf x},t)+\beta p({\bf x},t)\nabla \phi({\bf x})\right)
=-Hp({\bf x},t).
\end{equation}
Here $\phi$ is a potential which is periodic and finite, so the  diffusion constant at late times exists. In any dimension the late time effective diffusion constant is given by \cite{derr,scat}
\begin{equation}
\kappa_e = \kappa - {\kappa^2\beta^2 \over d} \int d{\bf x} d{\bf x}_0
\nabla \phi({\bf x}) \cdot H^{-1}({\bf x}, {\bf x}_0) \nabla \phi({\bf
x}_0) p_{eq}({\bf x}_0).
\end{equation}
This result can be demonstrated by adapting a method introduced by Derrida \cite{derro} for  discrete random walks to the case of  Langevin processes \cite{derr} or by direct manipulation of the  Langevin equation \cite{scat} to obtain a Kubo-like formula for the late time diffusion constant. However, if we consider a periodic potential in a region of space much larger than the period length and assume that the particle is in equilibrium (specifically  we mean that the position of the particle modulo the period is in equilibrium), it can be shown, directly from the Langevin equation \cite{scat}, that 
\begin{equation}
\kappa(t) = \kappa_e  +{\kappa^2 \beta^2\over dt}\int d{\bf x} d{\bf x}_0 \nabla \phi({\bf x})[ H^{-2}(1-\exp(-t
H))]({\bf x}, {\bf x}_0)\cdot \nabla \phi({\bf x}_0) p_{eq}({\bf x}_0).\label{eftc}
\end{equation}
In the above, $p_{eq}({\bf x})$ is the equilibrium probability distribution over a region of large but finite size $L$.

To start our analysis of these formulas we start by considering the correction to the late time, effective, diffusion constant by the presence of a potential in one dimension. We write $\kappa_e= \kappa -\Delta \kappa$ 
where
\begin{equation}
\Delta\kappa = {\kappa^2\beta^2}\int dx dx_0 {d\phi(x)\over dx} H^{-1}(x,x_0) {d\phi(x_0)\over dx_0}p_{eq}(x_0),
\end{equation}
and where the equilibrium distribution over the interval $[0,L]$ is given by
\begin{equation}
p_{eq}(x) = {\exp(-\beta \phi(x))\over L \langle \exp(-\beta \phi)\rangle}.
\end{equation}
In the above we have defined the spatial average of the Boltzmann weight
\begin{equation}
\langle \exp(-\beta \phi)\rangle = {1\over L}\int^L_0 dx\exp\left(-\beta\phi(x)\right) .
\end{equation}
Now we define
\begin{equation}
f(x) = \int dx_0 H^{-1}(x,x_0) {d\phi(x_0)\over dx_0}p_{eq}(x_0)\label{fx},
\end{equation}
and thus $f$ obeys
\begin{equation}
Hf = - \kappa{d\over dx}\left( {df\over dx} +\beta {d\phi\over dx}  f\right)= {d\phi(x)\over dx}p_{eq}(x_0).
\end{equation}
This second order differential equation can be integrated  to give
\begin{equation}
f(x)= {1\over \kappa\beta L\langle \exp(-\beta \phi)\rangle}\exp\left(-\beta \phi(x)\right)\left[ x\
+ c \int_0^x dx'\exp\left(\beta \phi(x')\right) + b\right],\label{solf}
\end{equation}
where $c$ and $b$ are two integration constants that must be determined. To determine the constants  we now consider a periodic potential with period $l$ such that $l\ll L$. The function $f$ must therefore be periodic with the same period, and this gives
\begin{equation}
c= -{l\over \int_x^{x+l} dx'\exp\left(\beta \phi(x')\right)} = -{1\over \langle \exp(\beta \phi)\rangle}.
\end{equation}
At this point we do no have to identify the constant $b$ as we have
\begin{equation}
\Delta \kappa = {\kappa^2\beta^2}\int dx {d\phi(x)\over dx} f(x),
\end{equation}
and  the term  proportional to $b$ in  $f(x)$ will contribute $0$ to $\Delta\kappa$
due to the periodicity of the potential. Performing the integrals we find
\begin{equation}
\Delta \kappa = {\kappa^2\beta^2\over  \kappa\beta L\langle \exp(-\beta \phi)\rangle}
{L\over \beta}\left(\langle \exp(-\beta\phi)\rangle - {1\over \langle \exp(\beta\phi)\rangle}\right),
\end{equation}
which simplifies to give
\begin{equation}
\Delta\kappa = \kappa - {\kappa\over \langle \exp(\beta\phi)\rangle\langle \exp(-\beta\phi)\rangle},
\end{equation}
and, eventually,
\begin{equation}
\label{kap}
\kappa_e = {\kappa\over \langle \exp(\beta\phi)\rangle\langle \exp(-\beta\phi)\rangle}.
\end{equation}
This  is a standard result on one dimensional diffusion which has been known for a long time and has been derived with a wide variety of different methods \cite{1d}. Interestingly the derivation we present here is based on the direct evaluation of the  MSD, whereas in all the other derivations we are aware of \cite{1d} a mean first passage time argument is used and then turned around to give the diffusion constant.

We now consider the finite time corrections. Our results on CTRWs suggest that 
the first term in the finite time integral  Eq. (\ref{eftc}) will be the leading correction, {\em i.e.} 
\begin{equation}
\kappa(t) \approx \kappa_e + {C\over t},
\end{equation}
where
\begin{equation}
C = {\kappa^2 \beta^2\over d}\int d{\bf x} d{\bf x}_0 \nabla \phi({\bf x})[ H^{-2}]({\bf x}, {\bf x}_0)\cdot \nabla \phi({\bf x}_0) p_{eq}({\bf x}_0).
\end{equation}
Staying again in one dimension, this can be written as
\begin{equation}
C = {\kappa^2 \beta^2}\int dx  g(x) f(x),
\end{equation}
where $f(x)$ is as defined by Eq. (\ref{fx}) and we have used the fact that $H^{-1\dagger} = H^{\dagger-1}$ (where $\dagger$ denotes the adjoint) to introduce
\begin{equation}
g(x) = \int dx_0 H^{\dagger-1}(x,x_0) {d\phi(x_0)\over dx_0}\label{gx}.
\end{equation}
This means that rather than solve a fourth order differential equation, we need only solve two second order ones. The first equation for $f$ is already solved in  Eq. (\ref{solf}) and the equation for $g$ is given by
\begin{equation}
-\kappa\left({d^2g\over dx^2} -\beta {d\phi\over dx }{dg\over dx }\right) = {d\phi\over dx}.
\end{equation}
This equation integrates to give
\begin{equation}
g(x) = {1\over \kappa \beta}\left[x -{1\over \langle\exp(\beta\phi)\rangle}\int_0^x dx' \ \exp\left (\beta\phi(x')\right) + b'\right],
\end{equation}
where the periodicity of $g$ has again been used to determine one of the constants of integration.

We thus see that the two functions $f$ and $g$ are related and we can write
\begin{eqnarray}
f(x) &=& {1\over \kappa\beta L\langle \exp(-\beta \phi)\rangle}\exp\left (-\beta\phi(x')\right) \left[ R(x)+ b\right] \nonumber \\
g(x) &=& {1\over \kappa \beta}\left[  R(x) + b'\right]
\end{eqnarray}
where
\begin{equation}
\label{hitting}
R(x) =L \left( \frac{x}{L}- {1\over L \langle \exp(\beta\phi)\rangle}\int_0^x dx' \ \exp\left (\beta\phi(x')\right)\right) .
\end{equation}
Before we proceed, it is interesting to note  that $R(x)$ in Eq. \eqref{hitting} has  a probabilistic interpretation. The first term in the brackets on the right-hand-side of Eq. \eqref{hitting} defines the so-called hitting probability in absence of an external potential (see, e.g. \cite{sid}) - the probability that a free diffusion on the interval $[0,L]$, starting at $x$, will first hit the point $x=L$ without ever hitting the left extremity of the interval. In turn, the second term on the right-hand-side of Eq. \eqref{hitting} is exactly the analogous hitting probability for diffusion in presence of the potential $-\phi(x)$ (see, e.g. \cite{sid2}). 

The problem we are now faced with is that of determining the constants $b$ and $b'$. First let us consider the general solution to an equation of the form Eq. (\ref{fx}) for the function $f(x)$. It can be rewritten up to a constant multiplicative factor, which we drop for notational convenience, as
\begin{equation}
f(x) = \int_0^\infty dt \int dx_0 [\exp(-tH)](x, x_0) {d\phi(x_0)\over dx_0}\exp\left(-\beta\phi(x_0)\right).
\end{equation}
The integral over $t$ will converge as the function ${d\phi(x_0)\over dx_0}\exp\left(-\beta\phi(x_0)\right)$ is orthogonal to the left eigenfunction (which is constant) of eigenvalue $0$ of the Fokker-Planck operator $H$. However 
we have
\begin{equation}
[\exp(-tH)](x, x_0)= p(x,x_0,t),
\end{equation}
where $p$ is the transition density for the process. Conservation of probability $\int dx \ p(x,x_0,t)=1$
then gives
\begin{equation}
\int dx f(x) = \int_0^\infty dt \int dx_0 \int dx p(x,x_0,t)  {d\phi(x_0)\over dx_0}\exp\left(-\beta\phi(x_0)\right)
= \int_0^\infty dt \int dx_0  {d\phi(x_0)\over dx_0}\exp\left(-\beta\phi(x_0)\right)=0
\end{equation}
where we have used periodic boundary conditions. This means that the constant $b'$ associated with the function $g$ does not contribute to the calculation and we have 
\begin{equation}
C = {1\over L\langle \exp(-\beta \phi)\rangle}\int dx  \exp\left (-\beta\phi(x)\right)[R^2(x) + bR(x)],
\end{equation}
where $b$ is determined by $\int dx f(x) = 0$ and thus
\begin{equation}
b = -{\int dx \exp\left (-\beta\phi(x)\right) R(x)\over L\langle \exp(-\beta\phi)\rangle}.
\end{equation}
Now if we denote averages with respect to the  equilibrium Gibbs-Boltzmann measure via
\begin{equation}
\langle A(x)\rangle_{eq} ={ \int_0^Ldx \  \exp\left(-\beta\phi(x)\right) A(x)  \over L\langle \exp(-\beta\phi)\rangle},
\end{equation}
we can write
\begin{equation}
\label{C}
C = \langle R^2(x)\rangle_{eq} -  \langle R(x)\rangle^2_{eq}.
\end{equation}
Note also that due to the periodicity of $\phi$ and $R$ with respect to $l$ the above averages can be computed over an interval of length $l$. An interesting consequence of this result is that the value of $C$ is always positive, in contrast to the CTRW case. Another interesting feature of this formula for $C$ is that it is independent of $\kappa$. 

Now consider the case where the potential is given by
\begin{equation}
\phi(x) = V({2\pi x\over\l}), \label{eqcs}
\end{equation}
where $V$ has period $2\pi$ giving $\phi$ a period $l$. Making the change of variables $z= 2\pi x/l$ in the integral expressions for $C$ we have
\begin{equation}
\label{R}
 \langle R^2(x)\rangle_{eq} = {l^2\over (2\pi)^2\int_0^{2\pi} dz'\exp\left(-\beta V(z')\right) }\int_0^{2\pi}dz\  \exp\left(-\beta V(z)\right) \left[ z -{\int 
_0^z dz' \exp\left(\beta V(z')\right)\over {1\over 2\pi}\int_0^{2\pi} dz'\exp\left(\beta V(z')\right) }\right]^2.
\end{equation}
and 
\begin{equation}
\label{Ra}
 \langle R(x)\rangle_{eq} = {l\over (2\pi)\int_0^{2\pi} dz'\exp\left(-\beta V(z')\right) }\int_0^{2\pi}dz\  \exp\left(-\beta V(z)\right) \left[ z -{\int 
_0^z dz' \exp\left(\beta V(z')\right)\over {1\over 2\pi}\int_0^{2\pi} dz'\exp\left(\beta V(z')\right) }\right].
\end{equation}
This means that $C = c l^2$, where $c$ is a constant independent of $l$. In the limit $\beta\to\infty$, for sufficiently smooth potentials, both 
$\kappa_e$ and $C$ can be evaluated the saddle point method. The effective diffusion constant
takes the Arrhenius, and more precisely Kramers \cite{kramers}, form 
\begin{equation}
\kappa_e= \kappa 2\pi \beta \sqrt{|V''(z_{\rm max})| V''(z_{\rm min})}
\exp\left(-\beta(V(z_{\rm max})-V(z_{\rm min})\right),
 \end{equation}
where $z_{\rm max}$ and $z_{\rm min}$ are respectively the points where the potential takes its maximal and minimum values respectively.
In the same regime the constant $C$ is given by
\begin{equation}
C = {l^2\over (2\pi)^2\beta V''(z_{\rm min})}.
\end{equation}

We therefore see that the diffusion constant at low temperatures is dominated by an Arrhenius like term involving the maximal energy barrier in the system, while the prefactors depends on the curvature of the maximum and minimum of the potential. The constant term $C$ however only depends on the  the curvature of the minimum of the potential.  
  
\begin{figure}[t]
   \centering
  \includegraphics[width=12cm]{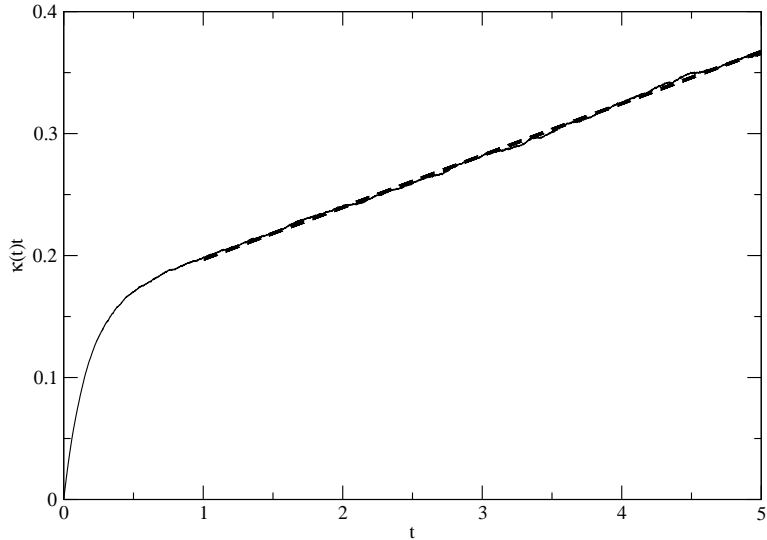}
  \caption{ $\kappa(t)t$ estimated from the MSD in a  numerical simulation of $10^5$ Langevin particles in the  potential in Eq. (\ref{eqcs}) given by $V(x)=\cos(x)$, $\beta=3$ and $l=4.0$ (continuous) black line. Shown as thick dashed line is the linear fit of $\kappa(t)t$ for $t>1$ yielding the estimate $\kappa_e=0.04258$ and $C= 0.15432$. The analytical predictions are $\kappa_e=0.04198$ and $C= 0.15778$.  }
  \label{msd}
\end{figure}

The predictions can be tested by simulating the Langevin equation. The particles are initially started at the same point and are then allowed to diffuse during an equilibriation time where the variable $x$ modulo the period $l$ can equilibriate. The initial conditions $X(0)$ for each particle used for the computation of the MSD $\langle (X(t)-X(0))^2\rangle  = 2\kappa(t)t$  are given by this equilibriation step. Two sources of error are present in the simulation (i) the use of a finite time step $dt$ and also the statistical fluctuations due to using a finite number of particles in the simulations. 
In the cases where $\kappa_e$ and $C$ are sufficiently large, that is to say larger than $0.1$ for $\kappa_e$ and $0.1$ for $C$,  the fluctuations (the difference in the fits for $\kappa_e$ and $C$ between two distinct simulations for the same number of particles but different random seed for the simulations) are of the order of $5\%$ of the measured values if we take $10^5$ particles. The time step is taken to be $dt = 0.0005$ (we checked that the results did not differ beyond the $5\%$ error in going between $dt=0.001$ and $dt=0.0005$). The MSD at each time is computed by ensemble averaging over each individual particle's squared displacement. From the MSD we plot the resulting estimate for $\kappa(t)t$. An example is shown in Fig (\ref{msd}). We see that after a certain time the MSD curve becomes linear, in this case for $t>1$. From the linear fit in this region the resulting estimates for  $\kappa_e$ and $C$ are extracted and we find that they are in excellent agreement with our analytical predictions.

The explicit calculations carried out here are of course relevant to one dimension. However from the  general form of Eq. (\ref{eftc}) we see that in any dimension if the constant $C$ exists then by dimensional analysis we must have that $C= cl^2$, where $c$ is independent of the periodicity $l$ of the potential. In addition,  using the explicit form of $H$, we see that the constant $C$ must be independent of $\kappa$.

\begin{figure}[t]
   \centering
  \includegraphics[width=12cm]{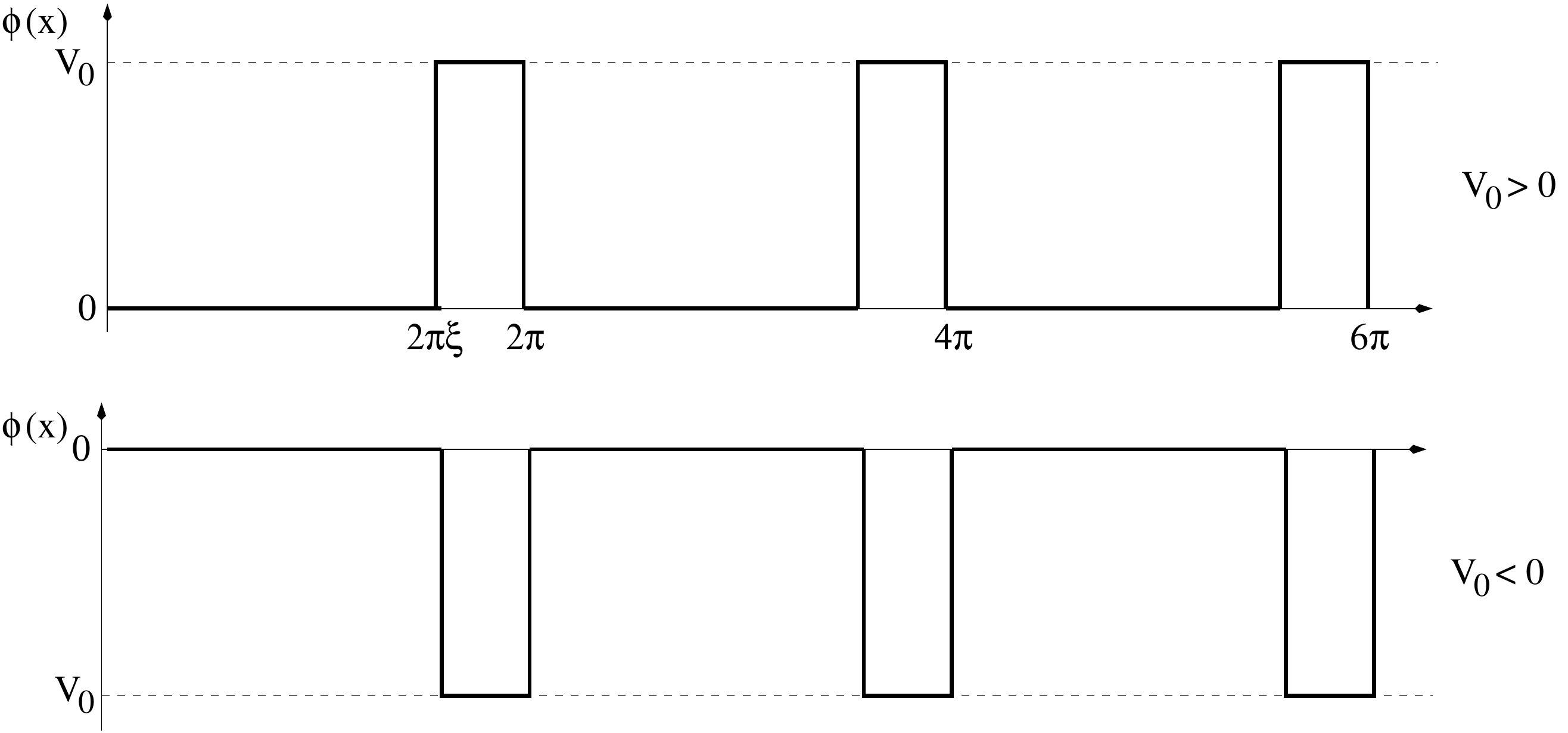}
  \caption{ A homogeneous potential $\phi(x) = 0$ perturbed by a narrow barrier with height $V_0 > 0$ or a well of depth $V_0 < 0$.}
  \label{2}
\end{figure}

Closed form expressions are not available for the constant $C$ for arbitrary potentials,
However, exploiting the exact results in Eqs.(\ref{C}), (\ref{R}) and (\ref{Ra}) for the amplitude $C$ of the relaxation term, we consider a simple, but instructive,  case   (see Fig.(\ref{2})) where the potential $\phi(x)$ is homogeneous, $\phi(x)=0$, for $0 \leq x < 2 \pi \xi$ with $0< \xi < 1$, and is perturbed by a narrow rectangular barrier (well), $\phi(x) = V_0$, in the region $2 \pi \xi \leq x \leq 2 \pi$, where $V_0 > 0$ in the case of a barrier and $V_0 < 0$ for a well.  
For this case  it is straightforward to find from Eq.(\ref{kap}) that the diffusion coefficient is given explicitly by
\begin{eqnarray}
\kappa_e &=& \frac{\kappa l^2}{(2 \pi)^2 \left(\xi + (1 - \xi) \exp(\beta V_0)\right) (\left(\xi + (1 - \xi) \exp(-\beta V_0)\right)} \nonumber\\
&=& \frac{\kappa l^2}{(2 \pi)^2 \left(\xi^2 + (1 - \xi)^2 + 2 \xi ( 1 - \xi) \cosh(\beta V_0)\right)}
\end{eqnarray}
Note that  $\kappa_e$ is an even function of $V_0$ which means that  the diffusion coefficient $\kappa_e$ is the same for both the case of periodic barriers and for the case of the wells with the same, by absolute value, $V_0$. Therefore, knowing just the diffusion coefficient $\kappa_e$, we are unable to distinguish if the diffusion takes place in presence of barriers or in the presence wells - even if we know $\xi$, upon extracting $\kappa_e$ from the numerical data we are only able to infer the absolute value of $V_0$  but not its sign. 
\begin{figure}[t]
   \centering
  \includegraphics[width=12cm]{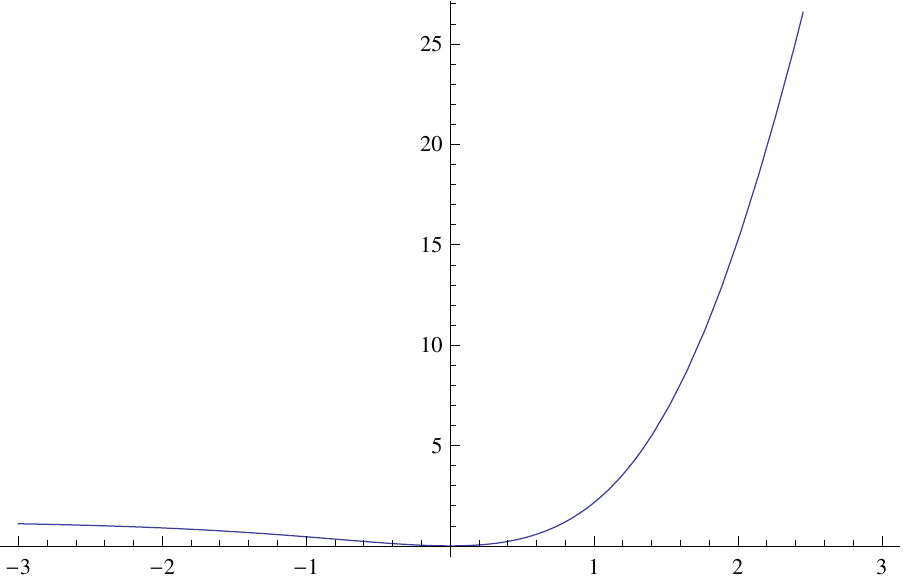}
  \caption{ The scaled relaxation amplitude $12 C/\xi^2 (1-\xi)^2 l^2$ in \eqref{barr} versus $\beta V_0$ for $\xi = 0.9$.}
  \label{3}
\end{figure}
Consider next what information one can extract by studying the amplitude of the relaxation term. From our Eqs.(\ref{C}), (\ref{R}) and (\ref{Ra}) we obtain
\begin{equation}
\label{barr}
C = \frac{\xi^2 (1 - \xi)^2 l^2}{12} \left(\frac{\exp(\beta V_0) - 1}{\xi + (1 - \xi) \exp(\beta V_0)}\right)^2 \,.
\end{equation}
A salient feature of the result in \eqref{barr} is that $C$, in a striking contrast to the diffusion coefficient $\kappa_e$, is a strongly 
asymmetric function of $\beta V_0$, as one observes in Fig.(\ref{3}). Namely, one gets a very different values of $C$ for $V_0 < 0$ and $V_0 > 0$, so that studying the amplitude of the relaxation term we can distinguish between diffusion in presence of barriers or and diffusion in presence of wells.

\section{Conclusions}

In many experimental situations diffusion constants are determined by a linear fit of an experimentally generated  MSD curve. In general Brownian particles are subjected to external forces and this extra drift means that they do not behave as pure Brownian motion. The effect of interaction means that  the average MSD only becomes proportional to $t$ at late times. In this region the diffusion constant can be extracted via a fitting procedure, for instance a linear fit (which may not necessarily be the best
way to fit such data \cite{fits1,fits2}) will yield a slope proportional to $2\kappa_e$ and also a constant $2C$ where it intercepts the vertical axis. If the particle is Brownian in an external conservative force field, we have shown that this constant must be positive and in one dimension we have found an analytical expression for it. We believe that these results could be useful in analyzing single particle tracking data, as the constant term, which is usually ignored or not discussed, contains potentially interesting information about the potential landscape seen by the Brownian particle. In particular we have seen that at low temperatures, knowing $C$ enables one to estimate the curvature of the local minima of the potential responsible for slowing diffusion down, while in the Kramers expression for the diffusion constant  this information is mixed with the Arrhenius term which contains information about the energy barrier. 

A number of questions remain open, here we have examined the leading order correction to the average MSD which turns out to be a constant. The next order corrections would be interesting to analyze as they would allow one to estimate when one achieves the effective linear regime of asymptotic diffusion. Another important question is what happens in the case where the potential is not periodic but say, for instance, random but statistically stationary. In this case the leading order correction is probably time dependent \cite{scat}.

\end{document}